\documentclass[graybox]{svmult}

\usepackage{mathptmx}       % selects Times Roman as basic font
\usepackage{helvet}         % selects Helvetica as sans-serif font
\usepackage{courier}        % selects Courier as typewriter font
\usepackage{type1cm}        % activate if the above 3 fonts are
\usepackage{makeidx}         % allows index generation
\usepackage{graphicx}        % standard LaTeX graphics tool
\usepackage{multicol}        % used for the two-column index
\usepackage[bottom]{footmisc}% places footnotes at page bottom
\def\agile {\emph{AGILE}}

\def\comp {\emph{COMPTEL}}

\def\lum {\mbox{erg s$^{-1}$}}

\def\psr {PSR~B1509--58}

\makeindex             % used for the subject index
\begin{document}

\title*{\emph{AGILE} observations of 
PSR B1509-58: is QED photon splitting at work in pulsars?} 

\titlerunning{AGILE observations of PSR B1509-58}

\author{M.~Pilia and A.~Pellizzoni on behalf of the AGILE Team and Pulsar
Working Group}
\authorrunning{M.~Pilia and A.~Pellizzoni} 

\institute{M.~Pilia \at Dipartimento di Fisica, Universit\`a dell'Insubria,
  Via Valleggio 11, I-22100 Como, Italy, 
\email{mpilia@ca.astro.it}
\and A.~Pellizzoni \at INAF--Osservatorio Astronomico di Cagliari, localit\`a Poggio
  dei Pini, strada 54, I-09012 Capoterra, Italy \email{apellizz@ca.astro.it}}

\maketitle

\abstract*{We present the results of new \agile\ observations of \psr\ 
performed over a period of $\sim$2.5 years
following the detection obtained with preliminary data. 
The modulation significance of the lightcurve above 30 MeV is at a 
5$\sigma$ confidence level and the lightcurve is similar to those 
found earlier up to 30 MeV by \comp: a broad asymmetric first peak reaching 
its maximum $0.39 \pm 0.02$ cycles after the radio peak plus a second peak at
$0.94 \pm 0.03$.  
The $\gamma$-ray spectral energy distribution of pulsed flux is well
described by a power-law (photon index $\alpha=1.87\pm0.09$) with a remarkable 
cutoff at $E_c=81\pm 20$~MeV, representing the softest
spectrum observed among $\gamma$-ray pulsars so far.
The unusual soft break in the spectrum of PSR B1509-58 has been interpreted 
in the framework of polar cap models as a signature of the exotic photon
splitting process 
in the strong magnetic field of this pulsar. 
In the case of an outer-gap scenario, or the two pole caustic model, better
constraints on the geometry of the emission would be needed from the radio band in
order to establish whether the conditions required by the models to reproduce
\agile\ lightcurves and spectra match the polarization measurements. 
}

\section{Introduction}
\label{sec:1}
\psr\ was discovered as an X-ray pulsar with the {\it Einstein} satellite 
and soon  detected also at radio frequencies
(Manchester et al. 1982),  with  a  derived distance
supporting the  association with  the SNR MSH~15-52 ($d\sim 5.8$~kpc).  
With a period $P\simeq  150$~ms and a period derivative
$\dot P  \simeq 1.53 \times 10^{-12}$s~s$^{-1}$,  assuming the standard dipole
vacuum model, the estimated spin-down 
age for  this pulsar is 1570  years and  its inferred  surface
magnetic field is one of the  highest observed for an ordinary radio pulsar:
$B\approx 
3.1\times 10^{13}$~G, as calculated at the pole. Its rotational energy  loss
rate is $\dot E \approx 
1.8 \times  10^{37}$~erg/s. 

The young age  and the  high rotational  energy loss rate made this
pulsar a promising target for the $\gamma$-ray
satellites. In fact, the instruments on board of the
{\it Compton Gamma-Ray Observatory} ({\it CGRO}) observed its
pulsation at low $\gamma$-ray energies,
but it was not detected with high
significance by the {\it Energetic Gamma-Ray Experiment Telescope} ({\it EGRET}),
the instrument  operating at  the energies from 30~MeV to 30~GeV.

The Italian satellite \agile\ (Tavani et al. 2009) obtained the first detection of
  \psr\ in the {\it EGRET} band 
(Pellizzoni et al. 2009b)  confirming the
occurrence of a spectral break. 
Here we present the results of a $\sim 2.5$~yr monitoring campaign
of \psr\ with \agile, improving counts statistics, and therefore
lightcurve characterization, with respect to earlier \agile\ observations. 
With these observations the spectral energy distribution (SED) at
$E<300$~MeV is assessed, where the remarkable spectral turnover is observed.

\section{\emph{AGILE} Observations, Data Analysis and Results}
\label{sec:2}

\agile\ devoted a large amount of observing time to the region of \psr.
For details on \agile\ observing strategy, timing calibration and $\gamma$-ray
pulsars analysis the reader can refer to Pellizzoni et al. (2009a,b).
A total exposure of $3.8
\times 10^{9}$~cm$^2$~s ($E > 100$ MeV) was obtained during the $2.5$~yr
period of observations (July 2007 - October 2009) which,
combined with \agile\ effective area, gives our observations a good
photon harvest from this pulsar.

Simultaneous radio observations of \psr\ with the
Parkes radiotelescope in Australia are ongoing since the epoch of
\agile's launch.
Strong timing noise was present
and it was accounted for using the
$fitwaves$ technique developed in the framework of the TEMPO2 radio
timing software (Hobbs et al. 2004, 2006).   
Using the radio ephemeris provided by the {\it Parkes} telescope, 
we performed the folding of the $\gamma$-ray lightcurve including the wave
terms (Pellizzoni et al. 2009a).  
An optimized analysis followed, 
aimed at cross-checking and maximization 
of the significance of the detection, including an energy-dependent events 
extraction angle around source position based on the instrument
point-spread-function (PSF). 
The chi-squared ($\chi^2$)-test applied to the 10 
bin lightcurve at $E>30$ MeV  gave a detection significance of $\sigma = 4.8$.
The unbinned $Z_n^2$-test gave a significance of $\sigma = 5.0$ with $n=2$
harmonics.  
The difference between the radio and $\gamma$-ray ephemerides was 
$\Delta P_{radio,\gamma}=10^{-9}$~s, at a level lower than the error in
the parameter, showing perfect agreement
among radio and $\gamma$-ray ephemerides as
expected, further supporting our detection and \agile\ timing calibration.

\begin{figure}

\includegraphics[width=5cm]{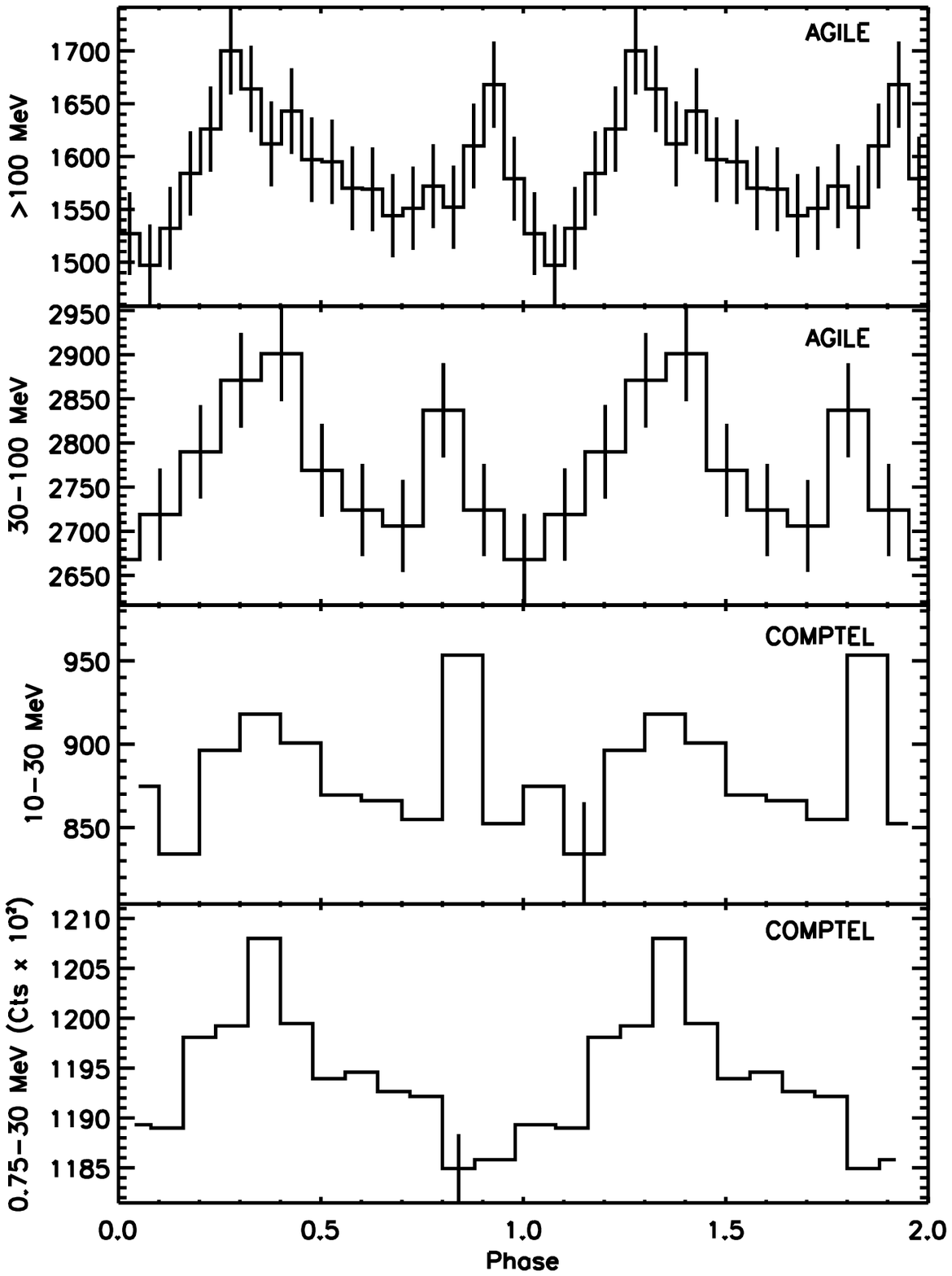} 
\hspace{\fill}
\includegraphics[width=6cm]{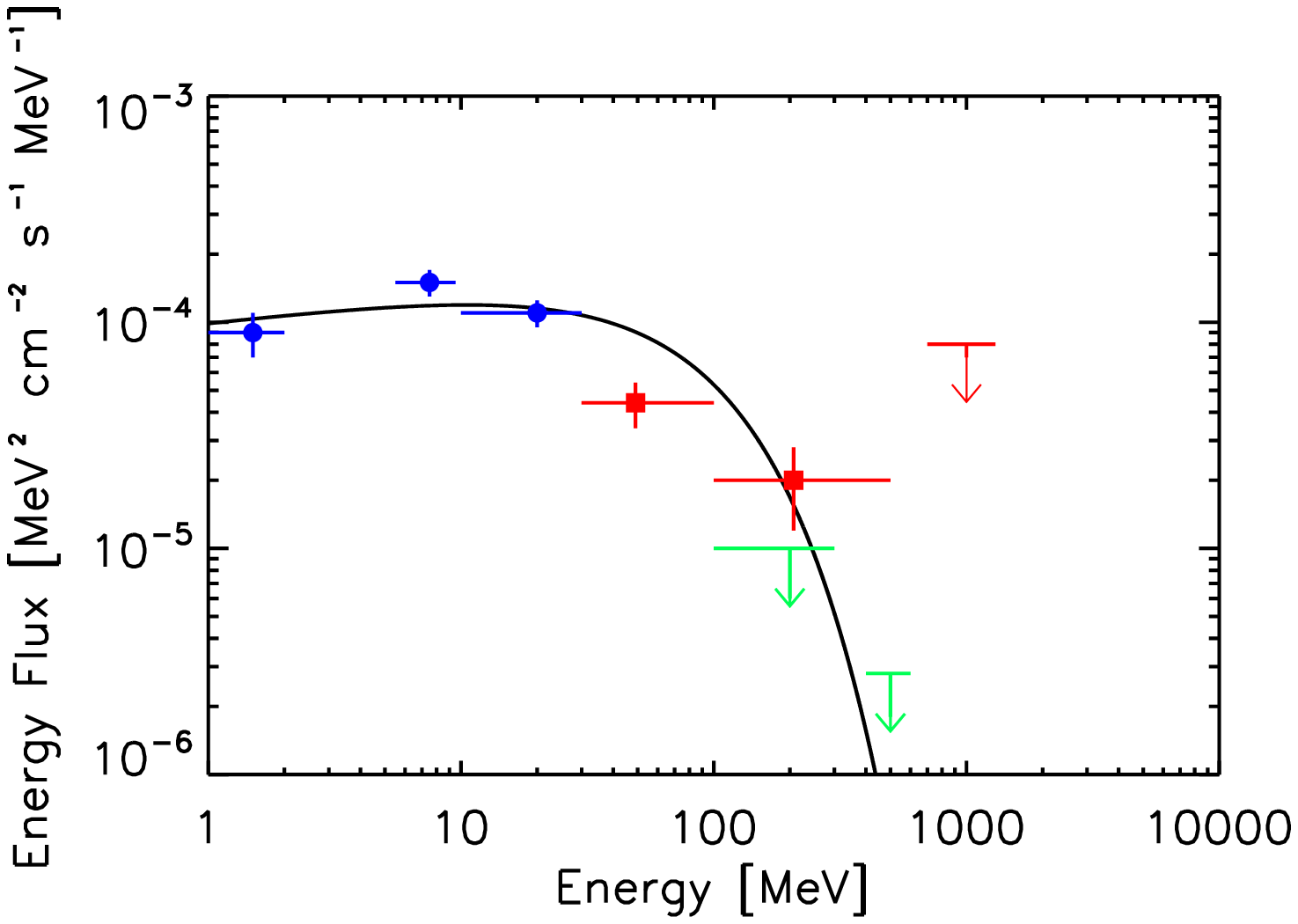}

\leftcaption{\label{fig:lc_tot}
Phase-aligned $\gamma$-ray light-curves of \psr\ with radio peak at
phase 0.
From the top: \agile\
 $>100$~MeV, 20 bins, 7.5
ms resolution; \agile\ $<100$~MeV, 10 bins, 15 ms
resolution; 
\comp\ 10--30~MeV and
\comp\ 0.75--30~MeV (from Kuiper et al. (1999).} 
\rightcaption{\label{spec} SED of \psr\ (solid
 line) obtained from a fit of pulsed fluxes from soft to hard
  $\gamma$-rays. The circular
   points represent \comp\ observations. The square points
  represent 
  \agile\ pulsed flux at $30<E<100$~MeV and $100<E<500$~MeV. 
The horizontal bar represents
  \agile\ upper limit above 500~MeV.  }  
\end{figure}

We observed \psr\ in three energy bands: 30--100~MeV, 100--500~MeV and above
500~MeV. 
We did not detect pulsed emission at a significance $\sigma \geq 2$ for $E >
500$~MeV. 
The $\gamma$-ray lightcurves of \psr\ for different energy bands
are shown in Fig. \ref{fig:lc_tot}. 
The \agile\ $E>30$ MeV lightcurve shows two peaks at phases
$\phi_1  = 0.39 \pm 0.02$ and $\phi_2  = 0.94 \pm 0.03 $ with respect to the
single radio peak, here put at phase 0.
The phases are calculated using a Gaussian fit to the peaks, yielding
a FWHM of $0.29(6)$ for the first peak and of $0.13(7)$ for the second peak,
where we quote in parentheses (here and throughout the paper) the 1$\sigma$ error
on the last digit.
The first peak is coincident in phase with \comp's peak (Kuiper et al. 1999). In its
highest energy band (10--30~MeV) \comp\ showed the indication of a second peak 
(even though the modulation had low significance, $2.1 \sigma$).
This second
peak is coincident in phase with \agile's second peak (Fig. \ref{fig:lc_tot}).
\agile\ thus confirms the previously marginal detection of a second peak.

Based on our exposure 
we derived the $\gamma$-ray flux from the number of
pulsed counts. 
The pulsed fluxes in the three \agile\ energy bands were 
$F_{\gamma}= 10(4)\times 10^{-7}
$~ph~cm$^{-2}$~s$^{-1}$ in the 30--100~MeV band, 
$F_{\gamma}= 2.1(5)\times 10^{-7} $~ph~cm$^{-2}$~s$^{-1}$ in the 100--500~MeV
band  
and a $1 \sigma$ upper limit $F_{\gamma}< 8\times 10^{-8}
$~ph~cm$^{-2}$~s$^{-1}$ for $E>500$~MeV.

Fig. 2 shows the SED of \psr\ based on
\agile's and \comp's observed fluxes. 
\comp\ observations suggested a 
spectral break between 10 and 30 MeV. 
\agile\ pulsed flux at energies $E > 30$~MeV
confirms the presence of a soft spectral break, but  
the detection of significant emission at $E>100$~MeV
hints to a cutoff at slightly higher energies.
As shown in Fig. \ref{spec}, we modeled
the observed fluxes with a power-law plus cutoff fit 
using the Minuit minimization package (James et al. 1975): $F(E)=k \times
E^{-\alpha}\exp[-(E/E_{c})^{\beta}]$, 
with three free parameters: the normalization $k$, the spectral index
$\alpha$, the cutoff energy $E_c$ and allowing $\beta$ to assume values of 1
and 2. 
No acceptable $\chi^2$ values were obtained for $\beta=2$,
while for an $\beta=1$ we found $\chi^2_{\nu}=3.2$ for $\nu = 2$
degrees of freedom, corresponding to a
null hypothesis probability of 0.05. 
The best values thus obtained for the parameters of the fit were:
$k=1.0(2)\times 10^{-4}$, $\alpha=1.87(9)$, $E_{c}=81(20)$~MeV.

\section{Discussion}
\label{sec:3}

The bulk of the spin-powered pulsar flux is usually emitted in the MeV-GeV
energy band with  
spectral breaks at $\leq 10$~GeV (e.g. Abdo et al. 2010).
\psr\ has the softest spectrum observed among $\gamma$-ray
pulsars, with a sub-GeV cutoff at $E \approx 80$~MeV. 

When \psr\ was detected in soft $\gamma$-rays but not significantly at $E>30$~MeV,
it was proposed that the mechanism 
responsible for this low-energy spectral break might be photon splitting
(Harding et al. 1997).
The photon splitting (Adler et al. 1970) is an exotic third-order quantum
electro-dynamics 
process expected when the 
magnetic field approaches or exceeds the $critical$ value defined as
$B_{cr}=m^2_e c^3/(e\hbar)=4.413\times 10^{13}$~G. 
In very
high magnetic fields the formation of pair cascades can be altered
by the process of photon splitting: $\gamma \rightarrow \gamma\gamma$. 

In the case of \psr\ a polar cap model with photon splitting would be
able to explain the soft $\gamma$-ray emission and the low energy
spectral cutoff, now quantified by \agile\ observations.
Based on the observed cutoff, which is related to the photons' saturation
escape energy, 
we can derive constraints on the magnetic field strength at emission,
in the framework of photon splitting:

\begin{equation}
\epsilon_{esc}^{sat} \simeq 0.077(B^{\prime}  \sin \theta_{kB,0})^{-6/5} 
\label{eq:emax}
\end{equation}

where $\epsilon_{esc}$ is the photon saturation escape energy,
$B^{\prime}=B/B_{cr}$ and $\theta_{kB,0} $ is the angle between the 
photon momentum and the magnetic field vectors at the surface and is here
assumed to be very small: 
 $\theta_{kB,0} \leq 0.57 ^{\circ} $
(Harding et al. 1997). 
Using the observed cutoff ($E= 80$~MeV) we find that $B^{\prime}
\geq 0.3$, which 
implies an emission altitude $\leq 1.3 R_{NS}$, which is the height where
also pair production could ensue.
This altitude of emission is in perfect agreement with the polar cap models
(see Daugherty \& Harding 1996). 
The scenario proposed by Harding et al. (1997) 
is strengthened by its prediction that PSR~B0656+14
should have a cutoff with an intermediate value between \psr\ and the
other $\gamma$-ray pulsars. 
Additionally, \psr\ (Kuiper et al. 1999, Crawford et al. 2001) and PSR~B0656+14 
(De Luca et al. 2005, Weltevrede et al. 2010) show
evidence of an aligned geometry, which could imply polar cap emission. 

The polar cap model 
as an emission mechanism is debated.
From the theoretical point of view, the angular momentum is
not conserved in polar cap emission (Cohen \& Treves 1972,
Holloway 1977, Treves et al. 2010).
And a preferential explanation of the observed $\gamma$-ray
lightcurves with high altitude cascades 
comes from the recent
results by {\it Fermi} (Abdo et al. 2010).
In the case of \psr, 
the derived $\gamma$-ray luminosity from the flux at $E > 1$~MeV,
considering a 1~sr beam sweep, is 
$L_{\gamma}=5.7^{+0.1}_{-0.5} \times 10^{35}$~erg/s.
The convertion efficiency of the rotational
energy loss ($\dot E \approx 1.8 \times  10^{37}$~\lum, see $\S$1) 
into $\gamma$-ray luminosity is 0.03.
If the $\gamma$-ray
luminosity cannot account for most of the rotational energy loss, then the
angular momentum conservation objection becomes less cogent for this pulsar.

Alternatively, 
an interpretation of \psr\ emission can be sought
 in the frame of the three dimensional outer gap model
 (Zhang \& Cheng 2000). 
According to their model, hard X-rays and low energy $\gamma$-rays are both
produced by synchrotron self-Compton radiation of secondary 
e$^+$e$^-$ pairs of the outer gap. Therefore, as observed, the phase
offset of hard X-rays and low energy $\gamma$-rays with respect to the radio
pulse is the same, with the possibility of a small lag due to the thickness of
the emission region. 
According to their estimates 
a magnetic inclination angle $\alpha\approx 60 ^o$ and a viewing
angle $\zeta \approx 75 ^o$ are
required to reproduce the observed lightcurve. 
Finally, using the simulations of Watters et al. 2009), 
the observed lightcurve from \agile\ is best reproduced
if $\alpha\approx 35 ^{\circ}$  and   $\zeta \approx 90
^{\circ}$, in the framework of the two pole caustic model
(Zyks \& Rudak 2003).

The values of $\alpha$ and $\zeta$ required by the Zhang \& Cheng model
are not in good 
agreement with the corresponding values obtained with radio measurements.
In fact, Crawford et al. (2001) observe that $\alpha$ must be $< 60 ^{\circ}$ 
at the $3 \sigma$ level.
The prediction obtained by the simulations of Watters et al. 
better agrees with the radio polarization 
observations.
In fact, Crawford et al. also propose that,
if the restriction is imposed that $\zeta > 70 ^{\circ}$ (Melatos 1997), 
then $\alpha > 30 ^{\circ}$ at the $3 \sigma$ level.
For these values, however, the Melatos model for the spin down of an oblique
rotator 
predicts a braking index $n>2.86$, slightly inconsistent with the observed 
value ($n=2.839(3)$).
Therefore, at present the geometry privileged by the
state of the art measurements is best compatible with polar cap models. 

\begin{acknowledgement}

M.P. thanks A. Treves for useful discussion and comments.
\agile\ is funded by the Italian Space Agency (ASI), with
programmatic participation by the Italian Institutes of Astrophysics (INAF) and
Nuclear Physics (INFN).
The Parkes radiotelescope is funded by the
Commonwealth Government as part of the ATNF, managed by CSIRO.

\end{acknowledgement}

\end{document}